\documentclass[mathleft,fleqn,%
]{an}
\usepackage{url}
\usepackage{graphicx}
\usepackage[varg]{txfonts}
\usepackage{natbib}
\bibpunct{(}{)}{;}{a}{}{,}
\begin{document}

% The following seven commands are intended for editorial usage and
% should be ignored by the author(s).
\Pagespan{1}{}% Document's page range. 
% If second parameter is left empty, the last page is computed
% automatically.
\Yearpublication{2016}%
\Yearsubmission{2016}%
\Month{0}%   
\Volume{999}%  
\Issue{0}% 
\DOI{asna.201400000}% 

\title{The CoRoT-GES Collaboration \\
      improving Red Giants spectroscopic surface gravity and abundances with asteroseismology}

\author{M. Valentini\inst{1}\fnmsep\thanks{Corresponding author:
        {mvalentini@aip.de}}
% Example for footnote, note the usage of the \texttt{fnmsep} command
% as separator between institute number and footnote mark}
\and C. Chiappini\inst{1}
\and A. Miglio\inst{2}
\and J. Montalb\'{a}n\inst{3,4}
\and T. Rodrigues\inst{3,4}
\and B. Mosser\inst{5}
\and F. Anders\inst{1}
\and the CoRoT RG group
\and the GES consortium 
}
\titlerunning{The CoRoT-GES Collaboration}
\authorrunning{M. Valentini}
\institute{
Leibnitz Institute f\"{u}r Astrophysics Potsdam (AIP), Potsdam, Germany
\and 
University of Birmingham, Birmingham, UK
\and
Dipartimento di Fisica e Astronomia, Universit\'{a} degli Studi di Padova, Padova, Italy
\and
INAF-Osservatorio Astronomico di Padova, Vicolo dell'Osservatorio 5, I-35122 Padova, Italy
\and
LESIA, Observatoire de Paris, PSL Research University, CNRS, Universit\'{e} Pierre et Marie Curie, Universit\'{e} Paris Diderot, 92195 Meudon, France }

\received{XXXX}
\accepted{XXXX}
\publonline{XXXX}
 

\abstract{%
Nowadays large spectroscopic surveys, like the Gaia-ESO Survey (GES), provide unique stellar databases for better investigating the formation and evolution of our Galaxy. Great attention must be paid to the accuracy of the basic stellar properties derived: large uncertainties in stellar parameters lead to large uncertainties in abundances, distances and ages. Asteroseismology has a key role in this context: when seismic information is combined with information derived from spectroscopic analysis, highly precise constraints on distances, masses, extinction and ages of Red Giants can be obtained. In the light of this promising joint-action, we started the CoRoT-GES collaboration. We present a set of 1,111 CoRoT stars, observed by GES from December 2011 to July 2014, these stars belong to the CoRoT field LRc01, pointing at the inner Galactic Disk. Among these stars, 534 have reliable global seismic parameters. By combining seismic informations and spectroscopy, we derived precise stellar parameters, ages, kinematic and orbital parameters and detailed element abundances for this sample of stars. We also show that, thanks to asteroseismology, we are able to obtain a higher precision than what can be achieved by the standard spectroscopic means. 
This sample of CoRoT Red Giants, spanning Galactocentric distances from 5 to 8 kpc and a wide age interval (1-13 Gyrs), provides us a representative sample for the inner disk population.}

\maketitle

\section{Introduction}
%The main questionT
Galactic Archeology, the study of how Milky Way formed and evolved, is nowadays entering in a golden era.
The forthcoming Gaia mission data releases \citep{Perryman2001} and the large spectroscopic Galactic surveys, like RAVE \citep{Steinmetz2006}, GES \citep{Gilmore2012}, APOGEE \citep{Majewski2015}, GALAH \citep{Freeman2010} and the future 4MOST \citep{deJong2012}, are offering wide, unique and promising stellar datasets for testing the modern chemo-dynamical models.

%What we have now
By comparing the main observables of the different components of our Galaxy with those predicted by models, like the age-metallicity relation, chemical gradients and kinematics, we will be able to understand the mechanisms that led to the actual Milky Way (i.e. \citet{Minchev2014}, and Minchev, Famay, Gerhard contributions to this conference). 
This comparison with models requires high precision and accuracy in distance, velocity, element abundances and ages.
Typically, for chemo-dynamical investigations, accuracies in velocity better than 1 km/s, few \% in distance and lower than 0.1 dex for element abundances are needed, in addition to an information on age, with an error lower than 20\%.

%Can these requirements being addressed today?
The accuracy in distance can be achieved thanks to Gaia, while the high precision on abundances can be addressed thanks to high resolution spectroscopy on high SNR spectra (e.g. 0.08 dex for GES and 0.05 dex for APOGEE). Most of the modern stellar spectroscopic surveys are targeting Red Giants, since they are the perfect tracers for Galactic investigations, thanks to their intrinsic brightness and incidence. However, for Red Giants stars, the atmospheric parameters determination from spectroscopy, especially for surface gravity, log(g), can be difficult. The surface gravity, in fact, can be affected by systematics up to 0.2 dex for this kind of stars (\citealp{Morel2012, Hekker2013, Heiter2015}). Since the abundances determination is coupled with atmospheric parameters, such systematics can lead to systematics of the same magnitude in the element abundances, compromising the quality of the data sample. Regarding the age determination, while it can be computed with a reasonable accuracy only for few targets (i.e. clusters) or dwarfs (with the standard method of isochrone fitting), it remains still precluded for field Red Giants: when using the commonly used isochrone fitting technique, age uncertainty can be up to 80\% \citep{Bergemann2014}, due to the degeneracies affecting the Red Giants locus.

%The question --> Asteroseismology is the answer
While waiting for Gaia data releases, asteroseismology can help in improving atmospheric parameters, abundances, age and distance for field Red Giants, required for Galactic Archaeology investigations.  
 
%Why asteroseismology 
The CoRoT and Kepler space missions revolutionised the view on Red Giants, showing that it is possible to directly link the two main seismic observables, $\Delta\nu$ and $\nu_{\rm{max}}$, to the stellar mass and radius. Thank to these scaling relations, it is therefore possible to determine a very precise and accurate log(g), with an error of only 0.03 dex (\citealp{Morel2012, Thygesen2012}). Fixing the gravity to the very precise log(g) provided by asteroseismology, abundances with a precision of 0.05 dex can be measured, as showed, for few stars, in \citet{Morel2014} and \citet{Batalha2011}. Since the seismic scaling relations provides a very precise value of the star's mass and radius (typical errors of 10\% and 3\% respectively), it is also possible to derive the stellar age (since the age of a Red Giant star is directly linked to its mass), even though always using models, and distance. Asteroseismology have been already successfully applied for better investigating disk population in \citet{Miglio2013} and in identifying a new population of young alpha-enhanced stars, see \citet{Chiappini2015} (CoRoT data) and \citet{Martig2015} (Kepler data). Nowadays asteroseismology has been included in the main spectroscopic surveys as a calibration tool,as in GES \citep{Pancino2012}, APOGEE \citep{Pinsonneault2014}, and LAMOST \citep{Wang2016}, where a benchmark of Red Giants possessing very good seismic parameters have been used for better testing, and eventually calibrating, the measured log(g), or as training set for their pipelines. 

%This paper with Corot
In this contribution we present how we analysed the spectra of the sample of CoRoT solar-like oscillating stars observed by the Gaia-ESO Survey (GES). These stars belong to the LRc01 field of CoRoT, pointing at the inner part of the Galactic disk. In section 2 we present how the sample of CoRoT Red Giants were selected and observed, in Section 3 we present how spectra have been analysed using asteroseismic information on gravity and how distances, ages, reddening and orbit parameters have been computed. Finally, in section 4, we present our conclusions. 
 
\section{The sample and observations}

%Figura 1-----------
 \begin{figure}
   \centering
   \includegraphics[width=\linewidth]{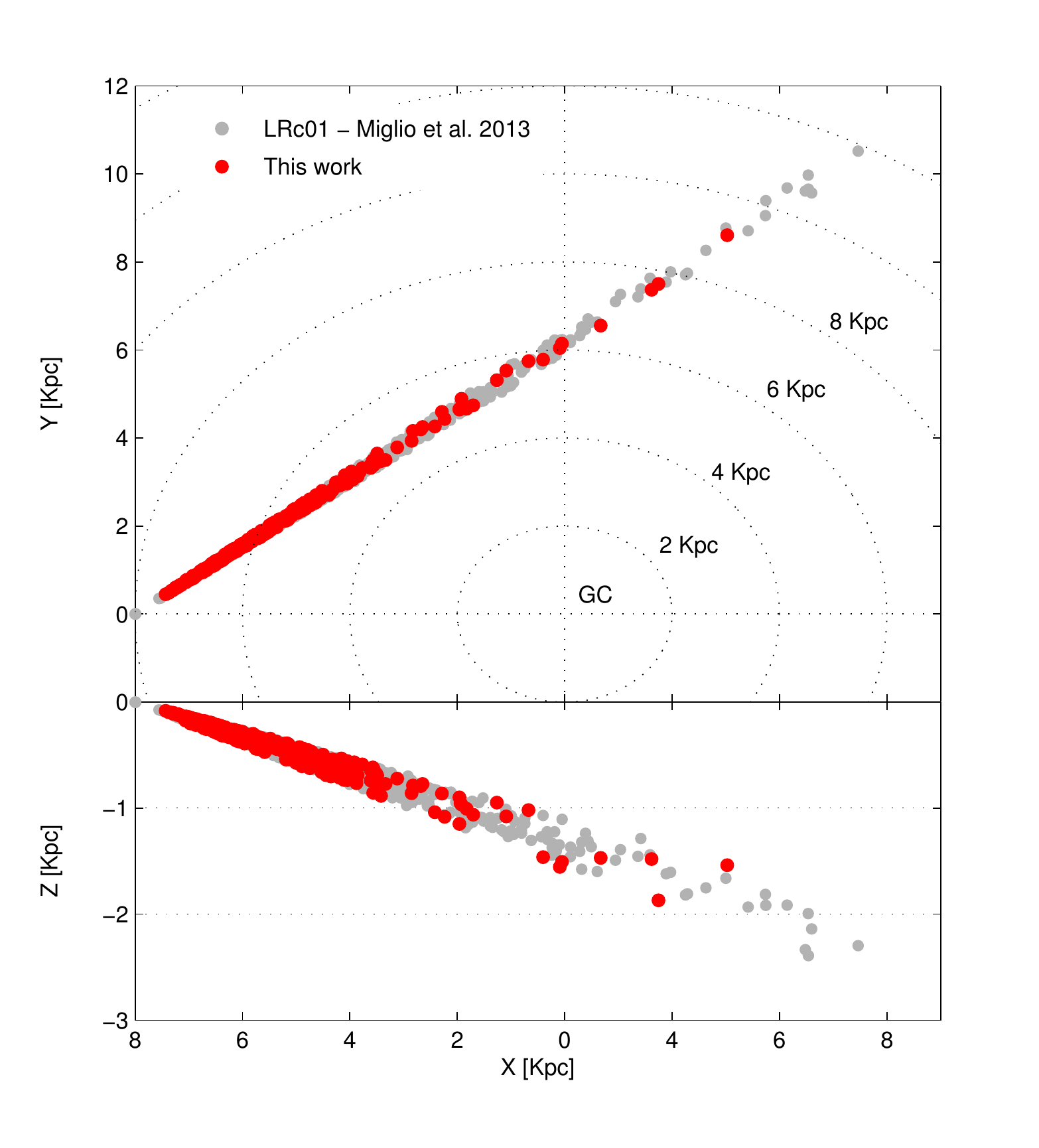}
	\caption{Spatial distribution of the CoRoT-LRc01 Solar-like oscillating Red Giants as in \citet{Miglio2013}. Stars analyzed in this work are color enhanced (red).}
    \end{figure}
%----------------------- 
%Figura 2-----------
 \begin{figure}
   \centering
   \includegraphics[width=\linewidth]{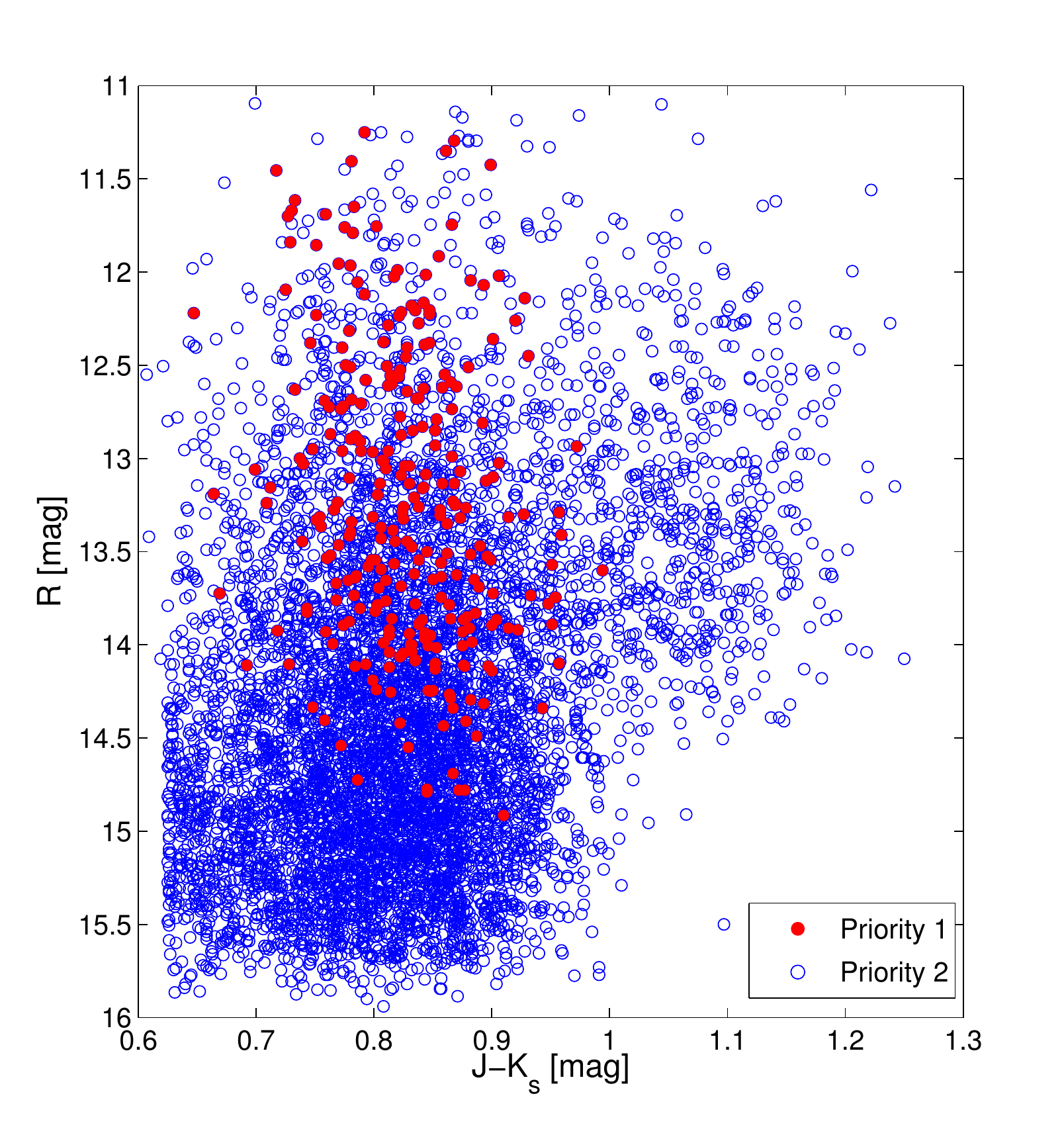}
	\caption{CMD of the 6845 objects in the CoRoT-GES target list. Targets belong to the LRc01 field of the CoRoT satellite. Priority 1 targets are stars possessing evolutionary status, following \citet{Mosser2011}.}
    \end{figure}
%----------------------- 

% Target selection
CoRoT LR fields are 1$^{\circ}$ x 2.5$^{\circ}$ wide, while the field of view of the ESO-FLAMES instrument, the one used by GES, is only 25 arcmin wide. This means that, for optimising the efficiency of the spectroscopic follow-up of the CoRoT Red Giants, we needed to chose the CoRoT field with the highest target density and prepare 32 fields. 
  
Following \citet{Mosser2010} and \citet{Mosser2011}, we selected the LRc01 field (centred at RA 19h 26m DEC +0$,^{\circ}$). LRc01 field has the highest target density of solar-like oscillating red giants among the CoRoT fields. It contains 1379 red-giants with detected solar like oscillations, homogeneously distributed on the area with a mean density of 43 targets per FLAMES pointing. In Fig.~1 the spatial distribution of the LRc01 targets with detected oscillations is showed. In this figure we adopted the distances of \citet{Miglio2013}. Stars belonging to the LRc01 field are distributed along a pencil beam, pointing at $\sim$30$^{\circ}$ from Galactic Center, and $\sim$20$^{\circ}$ below the Galactic Plane.  

In order to use all the 113 science fibres of the instrument, the CoRoT target list had been complemented with 5466 fainter CoRoT targets, photometrically selected as red giants (0.6$\leq$(J-K$_s$)$\leq$1.3 mag and R$\leq$16 mag) but with no seismic measurement available.
 
The final list of 6845 targets was divided in 2 priority groups:
   \begin{itemize}
      \item Priority 1: 283 Red Giants with $\Delta{\rm \nu}$, $\nu_{\rm max}$ and information on evolutionary status available (from Mosser et al.2011). High resolution (UVES) spectroscopy preferred.
      \item Priority 2: 6562 Red Giants candidates, some possessing $\Delta{\rm \nu}$ and $\nu_{\rm max}$, intermediate resolution spectroscopy (GIRAFFE) preferred.
   \end{itemize}

The Color-Magnitude-Diagram (CMD) of the CoRoT-GES input catalogue is showed in Fig.~2, Priority 1 targets are coloured in red, while Priority 2 targets are coloured in blue. For some of the brightest stars, with better quality light curves, seismic parameters have been used for inferring their evolutionary status (Clump vs RGB).
% Observations and datareduction
GES collects spectra using the ESO-FLAMES facility mounted at at the Paranal Observatory (Chile). It is a multi fiber instrument that allows observations using simultaneously two spectrographs: UVES (high resolution, R=47,000) and GIRAFFE (low resolution, R$\approx$19,000). We requested UVES setup U590 and GIRAFFE set-ups HR10, HR21, HR15b, in order to measure, when possible, abundances of several elements, in addition to the atmospheric parameters (T$_{\rm{eff}}$, log(g) and v$_{\rm{mic}}$): alpha-elements (O, Mg, Al, Si, Ca, Ti), Iron-Peak Elements (Sc, V, Cr, Mn, Ni, Fe), n-capture ones (Y, Sr, Zr), plus Na, Li and K.
      
In the period corresponding to Data Release 4, December 2011-July 2014, GES observed 1111 CoRoT objects of the target list we provided. Observation are summarised in the first part of Tab.~1. We originally requested high SNR spectra, minimum 100, but the final spectra have, on average, a lower SNR, $\sim$50.

All the spectra we analysed in this work have been reduced, calibrated and normalized by the GES consortium, see \citet{Sacco2014} and \citet{Smiljanic2014}. 

%Table targets -----------------------------------------------------------------
\begin{table}
\caption{Summary of the number of targets analysed in this work, following  different steps of the analysis.}             
\centering                          
\begin{tabular}{l c c}       
\hline\hline                
\textit{Observed} & UVES & GIRAFFE \\    
\hline                       
Priority    1 & 26 & 41 \\ 
Priority    2 & 12 &1032 \\
\hline                                 
 Tot.         & 38 & 1073 \\
\hline \hline
\multicolumn{3}{l}{\textit{Step 1: Seismology quality check}} \\
\hline 
Priority    1 & 23 & 40 \\ 
Priority    2 & 6 & 550  \\
\hline                                 
Tot.          & 29 & 590 \\
\hline \hline
\multicolumn{3}{l}{\textit{Step 2: SNR $>$ 18;  $|\log{(g)_{\rm seismo}}-\log{(g)_{\rm spectra}|}<$0.7}} \\
\hline  
Priority    1 & 17 & 39  \\ 
Priority    2 & 1 & 477 \\
\hline                                 
Tot.              & 18 & 516 \\
\hline \hline
\multicolumn{3}{l}{\textit{Step 3: PARAM code converged}} \\
\hline 
Priority    1 & 14 & 39 \\ 
Priority    2 & 1 & 455 \\
\hline                                 
Tot.              & 15 & 483 \\
\hline \hline
\multicolumn{3}{l}{\textit{Step 4: Orbit computation successful}} \\
\hline 
Priority    1 & 15 & 39 \\ 
Priority    2 & - & 254 \\
\hline                                 
Tot.              & 15 & 293 \\
\hline \hline
\end{tabular}
\label{Tab:obs}   
\end{table}
%-------------------------------------------------------------------------------

\section{Data analysis}
\subsection{Atmospheric parameters and abundances}
%Figura 3-----------
 \begin{figure}
   \centering
   \includegraphics[width=\linewidth]{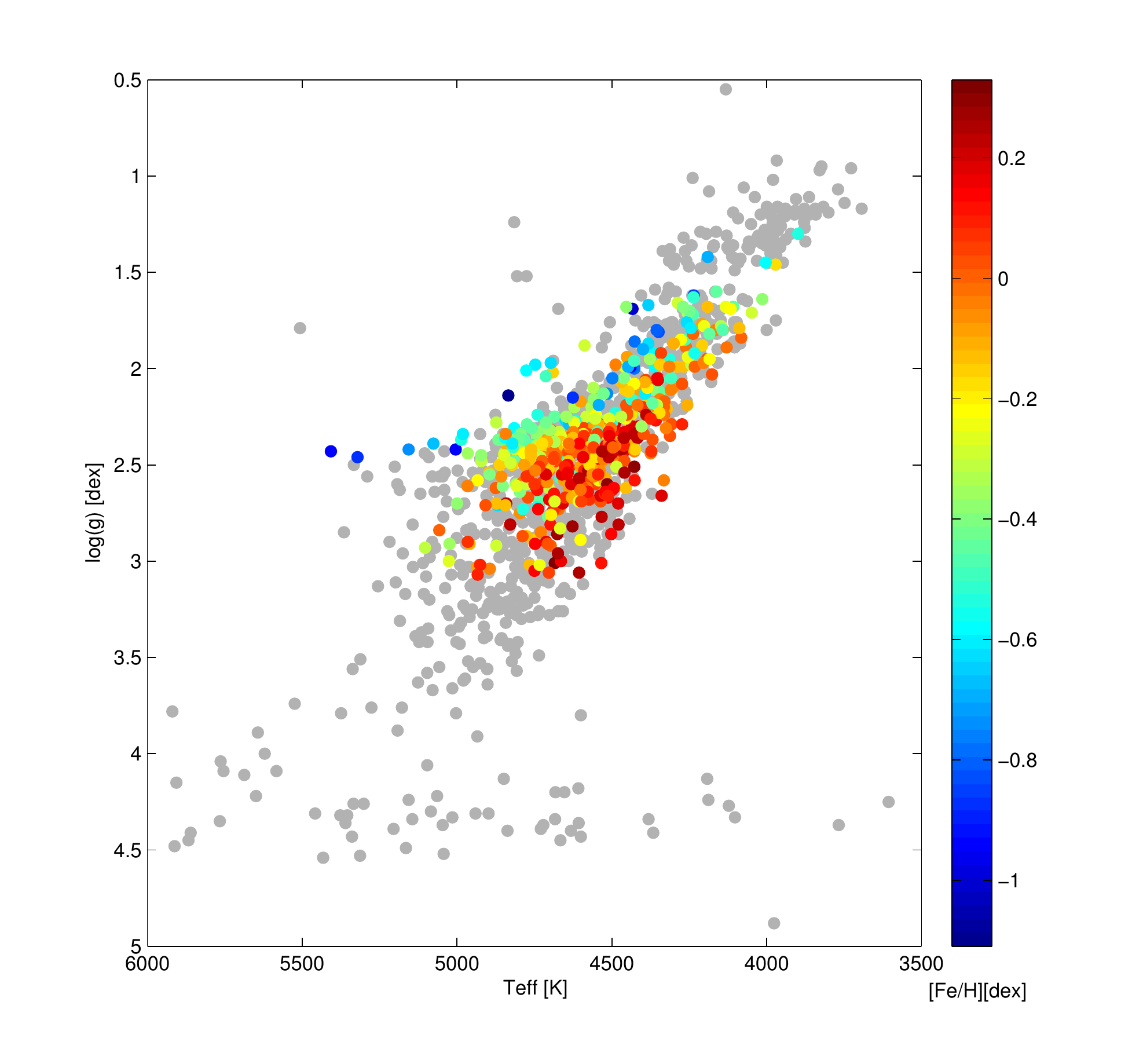}
	\caption{log(g)-$T_{\rm eff}$ distribution of the 1111 CoRoT-LRc01 Red Giants described in this work. Grey points are the object with atmospheric parameters derived without taking into account the seismic information, when available. Coloured dots (coloured following [Fe/H]) are object with atmospheric parameters derived by using seismic log(g).}
           \end{figure}
%----------------------- 

Thanks to the scaling relations widely discussed in this proceedings volume (e.g. Davies \& Miglio contribution), log(g) for Red Giants can be determined with high precision (typical error of 0.03 dex) and accuracy (seismic log(g) agrees with the one determined from the stellar mass and radius within ±0.06 dex), by using the seismic observable $\nu_{\rm max}$ and T$_{\rm eff}$:\\

$\log({\rm g})_{\rm seismo}=\log({\rm g})_{\odot}+\log\left(\frac{\nu_{\rm max}}{\nu_{\rm max\odot}}\right)+\frac{1}{2}\log\left(\frac{T_{\rm eff }}{T_{\rm eff\odot}}\right)$\\

Where $\nu_{\rm max\odot}$ = 3140.0 $\mu$Hz \citep{Pinsonneault2014}, T$_{\rm{eff}\odot}$ = 5777 K, log(g)$_\odot$ = 4.44 dex.

In this work we aim to introduce the use of the log(g)$_{\rm seismo}$ for iteratively deriving, using a spectroscopic pipeline (GAUFRE, \citet{Valentini et al. 2013}), more accurate atmospheric parameters and abundances for CoRoT-GES stars. In the past years this technique have been applied only on manual spectroscopic analysis (\citealp{Morel2014, Batalha2011, Mortier2014}), while only recently the seismic log(g) have been implemented in automatic pipelines (\citealp{Valentini2013, Hawkins2016}). 

Of the 1111 Corot targets observed by GES, 636 have detected solar like oscillations, following \citet{Mosser2010}. We therefore performed a reliability test on the seismic  $\Delta\nu$ and $\nu_{\rm max}$, using \citet{Bressan2012} isochrones. We compared the measured $\Delta\nu$ and $\nu_{\rm max}$ distribution of our targets with the one predicted by using isochrones (we used a set of 3 isochrones: [Fe/H]$=+$0.5 dex and age$=$1 Gyr, [Fe/H]$=0.0$ dex  and age$=$5 Gyr, and [Fe/H]$=-$1.0 dex and age$=$10 Gyr). By using the scaling relations we converted theoretical M and R into $\Delta\nu$ and $\nu_{\rm{max}}$, and then rejected those values not falling within 2$\sigma$ from the theoretical distribution. Following the isochrone test, 619 stars have reliable seismic $\Delta\nu$ and $\nu_{\rm max}$ (see Tab.~1, Step 1).

We therefore analysed spectra in two steps:
\begin{enumerate}
	\item We derived atmospheric parameters and abundances of the entire sample of observed CoRoT stars. High resolution spectra (UVES) were analysing using the classic Fe lines EW method ({\it GAUFRE-EW} module), while low resolution spectra were analysed using a $\chi^2$ technique on the synthetic library of de Laverny et al. (2012) ({\it GAUFRE-CHI2} module). Abundances were then derived using EW of lines and an ad-hoc model atmosphere (using MOOG {\it synthe} or {\it abfind} modules).
	\item For the 619 stars possessing seismic $\nu_{\rm max}$ we then derived atmospheric parameters iteratively, by fixing the log(g) to the seismic value (see Eq. 1), computed using the $T_{\rm eff}$ value coming from the latest iteration ({\it GAUFRE-EWseismo} and {\it GAUFRE-CHI2seismo} modules). On average, three iterations are needed for reaching log(g)-$T_{\rm eff}$ convergence. Abundances were then derived by using the same technique as in the previous step.
\end{enumerate}

During the spectroscopic analysis we performed a quality selection (Step~2 of Tab.~1). We considered only atmospheric parameters derived from spectra with SNR$>$18 (below this value spectra are too noisy for obtaining reliable values) and, when seismic informations are available, we considered those objects with a spectroscopically derived log(g) consistent with the seismic one: $\mid log(g)_{\rm seismo}-log(g)_{\rm spec}\mid\leq$ 0.7 dex. This last criterion was imposed for not forcing the pipeline to converge to a value too far from the original value, leading to unrealistic errors. This quality selection reduced our sample to 534 objects. The improvement on atmospheric parameters and abundances lead by the adoption of the seismic gravity is shown in Table~2.  

The log(g)-$T_{\rm eff}$~distribution of final sample of 534 CoRoT-GES Red Giants with good seismology and atmospheric parameters is plotted in Fig.~3 (over plotted to the original distribution obtained at step 1 of the analysis for all the 1111 targets). Stars are distributed within a log(g) interval of 1.3-3.1 dex, corresponding to the instrumental limits of the CoRoT satellite (a similar set of limits have been calculated for K-2 mission by \citet{Stello2015}. It is worth to notice that in the whole CoRoT-GES sample, with atmospheric parameters computed without taking asteroseismology into account (grey points in Fig.~3), many targets seems to be dwarfs (log(g)$>$3.5 dex). This dwarf contamination is due to a) a not optimal Priority 2 photometric colour selection, allowing the selection of some dwarf; b) wrong log(g) determination by the pipeline, due to bad SNR spectra or log(g)-T$_{\rm eff}$~degeneracies. 

\subsection{Atmospheric parameters validation}
A set of validation tests has been performed, using Gaia Benchmark stars (\citealp{Jofre2014, Heiter2015}) and CoRoGEE stars in common with GES \citep{Anders2016}. \\
{\bf GAIA BENCHMARK}

%TABLEliterature-----------------------------------------------------------
\begin{table}
      \caption[]{Mean difference and dispersions of atmospheric parameters for the 8 Gaia benchmark giant stars: atmospheric parameters measured with GAUFRE - literature values (\citealp{Heiter2015, Jofre2014, Jofre2015}).}
  \centering                          
\begin{tabular}{l c c c}       
\hline\hline                
 &         & UVES & UVES$_{fixed~log(g)}$ \\  
$<\Delta_{\rm T_{\rm eff}}>$ &[K] & 6 & -24 \\
$<\Delta_{\rm log(g)}>$ &[dex]& -0.10 & --  \\
$<\Delta_{\rm [Fe/H]}>$ &[dex]& -0.02 & 0.01  \\
$<\Delta_{\rm [Mg/H]}>$ &[dex]& -0.06 & -0.05 \\
 \hline \hline
  &         &  GIR & GIR$_{fixed~log(g)}$\\ 
$<\Delta_{\rm T_{\rm eff}}>$ &[K] &  -31 & -49 \\
$<\Delta_{\rm log(g)}>$ &[dex]&  -0.07 & -- \\
$<\Delta_{\rm [Fe/H]}>$ &[dex]&  -0.13 & -0.13 \\
$<\Delta_{\rm [Mg/H]}>$ &[dex]&  -0.01 & -0.06 \\
 \hline \hline
\end{tabular}
         \label{Tab:BenchCompT}
   \end{table} 
Gaia benchmark stars spectra were taken by GES with different setups and SNR. We analysed spectra taken with the same setup as CoRoT stars (GIRAFFE HR10+HR21 and UVES U580) and at the same SNR (50-100). We focused our test on the 8 giants (log(g)$<$3.5 dex) present in the sample. As reference values we used the log(g) and T$_{\rm eff}$ from \citet{Heiter2015}, while [Fe/H] and [Mg/Fe] values were taken from \citet{Jofre2014} and \citet{Jofre2015} respectively.

For simulating what happens by using the seismic log(g) we first analysed stars without any constraints, then we fixed the gravity to the literature value. As visible in Tab.~2, the agreement is good in both cases (offsets ~40 K in $T_{\rm eff}$, 0.01 dex in [Fe/H] and 0.05 dex in [Mg/H]), meaning that our pipeline provides reliable atmospheric parameters and abundances, even when fixing the gravity. In order to perform our investigation on a homogeneous set of data, we corrected the [Fe/H]$_{\rm GIRAFFE}$ of $+$0.13 dex, in order to have the Fe abundances obtained from GIRAFFE spectra on the same scale of the UVES Fe abundance (this correction will be adopted in the rest of the paper).

{\bf APOGEE}\\
%Figura field-----------
 \begin{figure}
   \centering
   \includegraphics[width=1\columnwidth]{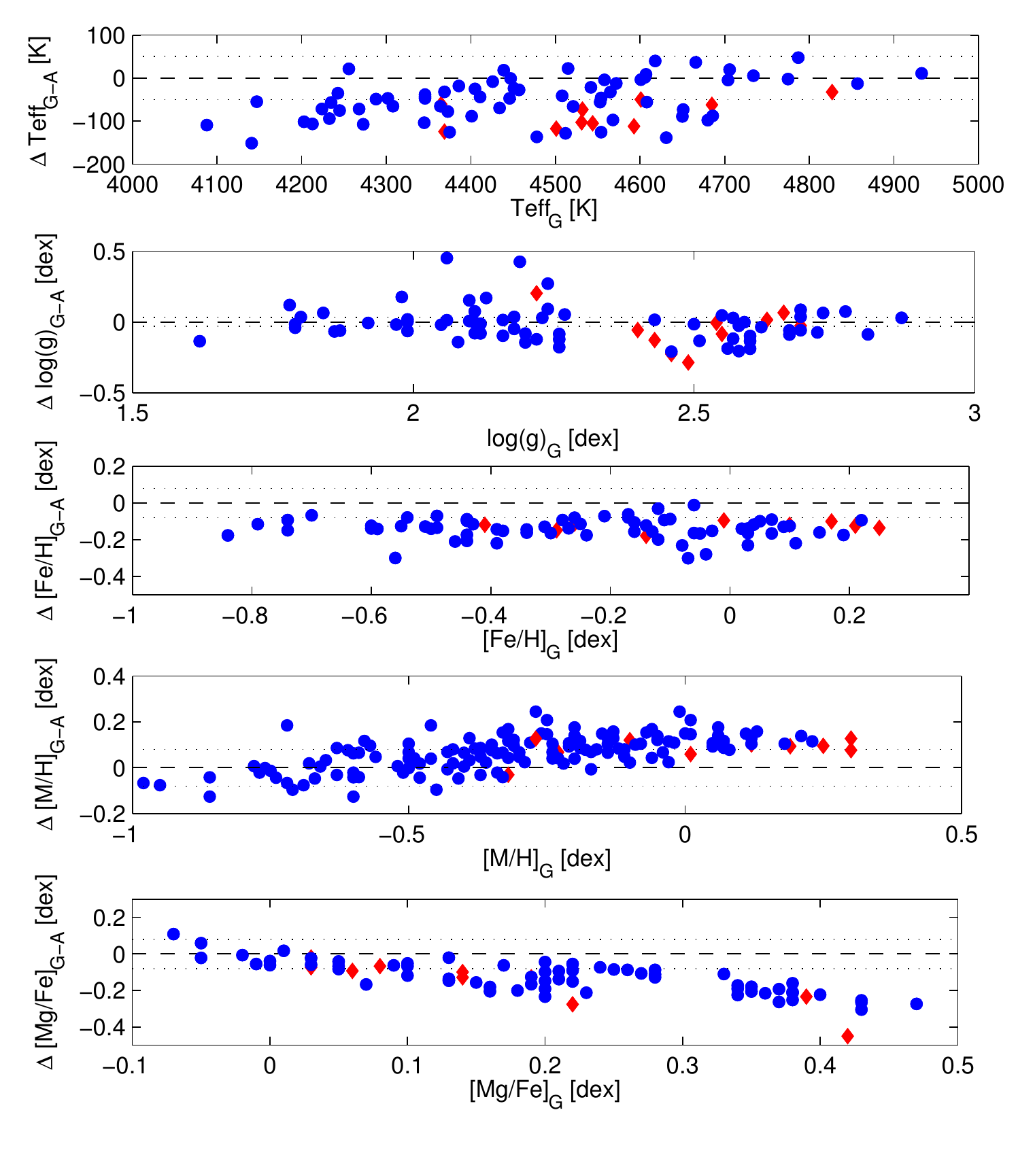}
	\caption{Atmospheric parameters difference between GAUFRE (G) and APOGEE-DR12 (A). From top to bottom: differences in $T_{\rm eff}$, log(g), [Fe/H], [M/H] and [Mg/H]. Values measured from UVES spectra are represented with red squares, values measured from GIRAFFE spectra are represented with blue circles ([Fe/H] is corrected). Dotted lines mark the mean $\pm 1 \sigma$ error to the atmospheric parameters derived by using GAUFRE.}
      \end{figure}
%------------------------
There are 77 CoRoT stars possessing asteroseismic values in common between the APOGEE survey (DR12) and this work. A detailed analysis of the CoRoT stars in APOGEE, CoRoGEE collaboration, has been performed in Anders et al. (2016).

We compared the atmospheric parameters derived by GAUFRE by adopting the log(g)$_{seismo}$ with the values taken from APOGEE-DR12 catalogue. The two surveys observe stars in different wavelength intervals (APOGEE in the infra-red, GES in the optical domain), and they analysed spectra using two different pipelines and linelists. It is worth to notice that CoRoGEE did not use seismic log(g) for refining the atmospheric parameters, but used a generic calibration based on seismic log(g)s from both Kepler and APOGEE. A comparison in atmospheric parameters and abundances ([Fe/H] and [Mg/Fe]) for the two surveys is shown in Fig.~4. There is a good agreement between the two surveys (no offset in log(g), a small offset of $\sim$75 K in $T_{\rm eff}$). An offset in [Fe/H] and [Mg/Fe] abundances of 0.1~dex has been measured, together with a small trend in [M/H] and [Mg/Fe], with APOGEE measuring slightly higher values. The investigation of these offsets and trends is beyond the purposes of this work, but they are probably the result of the small trend in T$_{\rm eff}$, and of the different pipelines and wavelength ranges adopted (a further discussion will appear in Valentini et al. (2016), in prep.) . 

\subsection{Distances, reddening, ages and orbit integration}
For each star possessing seismology we computed mass, radius, age, distance and reddening using the PARAM tool (da Silva et al. 2006, Rodrigues et al. 2014). As input information we adopted: the CoRoT $\Delta\nu$ and  $\nu_{\rm max}$ \citep{Mosser2010}, our refined atmospheric parameters, 2MASS \citep{Cutri2003} and WISE photometry \citep{Cutri2012}, and the information on the evolutionary status (available for 53 targets). PARAM converged for 498 objects (Step 3 of Tab.~1). On average, errors are: 7\% in mass, 3\% in radius, 21\% in age, 2\% in distance and 0.06 mag in Av. 

We then computed orbits, by using, when available, the proper motions of the UCAC4 catalogue. We, again, applied a quality selection on our sample, requiring that errors on pmRA and pmDEC do not exceed 80\%. Applying this quality criterion, we successfully computed orbits for 144 CoRoT-GES objects (Step 4 of Tab.~1), providing the orbital parameters: R$_{\rm apo}$ (apo-centric radius), R$_{\rm peri}$ (peri-centric radius), R$_{\rm mean}$ (mean radius), R$_{\rm guiding}$ (guiding radius), eccentricity $e$ and Z$_{\rm max}$ (maximum height from the Galactic plane). The implications for Galactic Archaeology investigations, about orbital parameters, ages and distances will be discussed in a forthcoming paper, Valentini et al. (2016) (in prep).
 
%%%%%%%%%%%%%%%%%%%%%%%%%%%%%%

\section{Conclusions}

%TABLE2--------------------------------------------------
\begin{table}
\caption{Typical errors on atmospheric parameters, abundances, mass, radius, age and distance obtained with classic techniques (e.g. spectroscopy, isochrone fitting) and the errors on the same values obtained using seismic information.}
\begin{tabular}{llcc} \hline \hline
$\sigma$  &       &  Spectroscopy & Spectroscopy + \\ 
          &       &               & Asteroseismology \\ \hline 
          &       &               &                  \\
$T_{\rm eff}$ & GIR. & 100 & 65 \\ 
$[\rm{K}]$  & UVES &  70 & 55 \\ 
 &         &             &                  \\
log(g) & GIR. & 0.20 & 0.03 \\ 
$[\rm{dex}]$            & UVES    & 0.12 & 0.03 \\
 &         &             &                  \\
$[\rm{Fe/H}]$ & GIR. & 0.10 & 0.08 \\ 
$[\rm{dex}]$          & UVES    & 0.09 & 0.05 \\ 
 &         &             &                  \\
$[\rm{elem./Fe}]$& GIR. & 0.20 & 0.08 \\
$[\rm{dex}]$         & UVES    & 0.08 & 0.05 \\ 
 &         &             &                  \\ \hline
$\sigma$  &       & & Asteroseismology \\ \hline 
          &       &               &                  \\
Mass            &         & - & 7\%  \\ 
Radius          &         &- & 3\%  \\
Age             &         & $>$80\%  & 21\% \\ 
Dist.           &         & - & 2\% \\ \hline \hline 
\end{tabular}
\end{table}
%---------------------------------------------------------

We derived, using a pipeline that implements seismic gravity in the analysis, refined atmospheric parameters and abundances of a sample of 534 CoRoT Red Giants observed by GES. In the analysis we fixed the log(g) to the seismic value, and we iteratively derived T$_{\rm eff}$~and overall metallicity [M/H], abundances of alpha-elements (O, Mg,Al, Si, Ca, Ti), n-capture elements (Y, Sr, Zr), Fe-peak elements (Sc, V, Cr, Mn, Ni, Fe), Na, Li and K. The typical errors on the atmospheric parameters and abundances, and the comparison with the errors obtained by using only spectroscopy, are reported on Tab.~3. The method was tested on the Gaia Benchmark stars and on a set of 77 CoRoT stars in common between GES and APOGEE, and resulting in reliable atmospheric parameters and abundances.

We finally obtained a sample of 498 stars, possessing not only precise abundances (typical error on element abundances $<$ 0.10 dex), but also distances and ages with an error of 2\% and 21\% respectively (see Table~3), in a more precise way than what can be obtained with the classic methods. Our sample is distribute along a beam pencil, spanning 5 - 8 Kpc in Galactocentric distance (see Fig.~1), and covering a wide age interval, from $\sim$1 Gyr to 12 Gyr. The use of this sample for Galactic Archaeology purposes is discussed in Valentini et al. (2016, in prep).

\acknowledgements
  The CoRoT space mission, launched on December 27 2006, was developed and operated
by CNES, with the contribution of Austria, Belgium, Brazil, ESA
(RSSD and Science Program), Germany and Spain. This research
has made use of the ExoDat Database, operated at LAM-OAMP,
Marseille, France, on behalf of the CoRoT/Exoplanet program.

\bibliographystyle{an}
\def\aj{AJ}\def\apj{ApJ}\def\apjl{ApJL}\def\araa{ARA\&A}\def\apss{Ap\&SS}
\def\mnras{MNRAS}\def\aap{A\&A}\def\nat{Nature}
\def\nar{New Astron. Rev.}
\bibliography{MValentini_2}

\end{document}